\documentclass[12pt,a4paper]{article}

\usepackage[utf8]{inputenc}
\usepackage{amsmath, amssymb, amsfonts}
\usepackage{graphicx}
\usepackage[a4paper, margin=1in]{geometry}
\usepackage{authblk}
\usepackage[colorlinks=true, linkcolor=blue, citecolor=blue, urlcolor=blue]{hyperref}
\usepackage{caption}
\usepackage{textgreek} 
\usepackage{bm} 
\usepackage{siunitx} 
\usepackage{booktabs} 
\usepackage{enumitem} 
\usepackage{ragged2e} 
\usepackage{csquotes} 
\usepackage[super]{natbib} 

\setcitestyle{authoryear=false,open={(},close={)},numbers,square} 

\usepackage{etoolbox}

\title{Non-invasive measurement of local stress inside soft materials with programmed shear waves}

\author[1]{Zhaoyi Zhang}
\author[2,*]{Guo-Yang Li}
\author[1]{Yuxuan Jiang}
\author[1]{Yang Zheng}
\author[3]{Artur L. Gower}
\author[4]{Michel Destrade}
\author[1,$\dagger$]{Yanping Cao}

\affil[1]{Institute of Biomechanics and Medical Engineering, AML, Department of Engineering Mechanics, Tsinghua University, Beijing 100084, PR China}
\affil[2]{Harvard Medical School and Wellman Center for Photomedicine, Massachusetts General Hospital, Boston, MA 02139, USA}
\affil[3]{Department of Mechanical Engineering, University of Sheffield, Sheffield, UK}
\affil[4]{School of Mathematical and Statistical Sciences, NUI Galway, Galway, Ireland}

\date{} 

\begin{document}

\maketitle

\begingroup
\renewcommand\thefootnote{}\footnote{*\href{mailto:gli26@mgh.harvard.edu}{gli26@mgh.harvard.edu}}
\footnotetext{$\dagger$\href{mailto:caoyanping@tsinghua.edu.cn}{caoyanping@tsinghua.edu.cn}}
\endgroup

\begin{abstract}
\justify
Mechanical stresses in soft materials across different length scales play a fundamental role in understanding the function of biological systems and in the use of artificial materials for engineering soft machines and biomedical devices. Yet it remains a great challenge to probe local mechanical stresses in situ in a non-invasive, non-destructive manner, in particular when the mechanical properties are unknown. To address this challenge, we propose an acoustoelastic imaging-based method to infer the local mechanical stresses in soft materials by measuring the speed of shear waves induced by custom-programmed acoustic radiation force. Using a medical ultrasound transducer to excite and track the shear waves remotely, we demonstrate the application of the method by imaging uniaxial stress and bending stress in an isotropic hydrogel, and the passive uniaxial stress in a skeletal muscle. These measurements were all done without the knowledge of the constitutive parameters of the materials. These examples indicate that our method will find broad applications, ranging from health monitoring of soft structures and machines, to the diagnosis of diseases that alter stresses in soft tissues.
\end{abstract}

\newpage

\section*{Introduction}
\justify
Mechanical stresses are important in biological and artificial soft materials across different length scales and play an essential role in their functions. For instance, adherent animal cells generate mechanical stress to migrate, divide, sense their environment, and communicate with other cells \citep{ref1, ref2, ref3, ref4}. At the tissue level, differential and/or constrained growth generates mechanical stresses that may trigger elastic instabilities and buckling patterns, leading to various morphological changes observed in nature \citep{ref5, ref6, ref7}. Forces produced by muscle contractions result in nearly all the movements in the human body \citep{ref8, ref9, ref10}. In short, it is fair to say that all living tissues are under mechanical stresses, even at rest, and understanding their distribution and magnitude is critical for uncovering the biophysics underpinning various life activities \citep{ref2}. Stresses play a vital role also in artificial soft materials \citep{ref11, ref12}, which are used, for example, in designing soft machines and developing wearable and implantable soft bioelectronics. Residual and/or applied mechanical stresses cannot be avoided in these applications \citep{ref10, ref13, ref14}. Being able to probe the mechanical stress in situ is needed for the optimal design of soft machines/instruments and for the evaluation of their mechanical behavior, e.g., fatigue life \citep{ref15, ref16}.

To date, it remains a great challenge to probe the mechanical stresses of soft materials in situ in a noninvasive manner, especially when their mechanical properties are not known \citep{ref2}. Traditionally, stresses can be inferred from measured deformations \citep{ref10, ref17} provided the mechanical properties and the undeformed configuration of the tested material are known. The hole-drilling method \citep{ref18, ref19} is such an example that enables the measurement of residual stress destructively. Many non-destructive methods have been developed, including ones that use X-rays, neutron diffraction, and ultrasonic waves \citep{ref19, ref20}, but these all require prior knowledge of the material constants or of the undeformed configurations of tested materials, all of which are challenging to acquire. For example, stress alters the speed of ultrasonic waves by the acoustoelastic effect \citep{ref19, ref21, ref22, ref23}. But its interpretation requires knowledge of the third-order elastic constants, and calibrating for these parameters is by no means trivial, even in controlled laboratory environments \citep{ref21, ref22, ref24, ref25}. Measuring the constitutive parameters of soft tissues in vivo or of artificial soft materials in service represents an even greater challenge. Moreover, the mechanical properties of these materials may vary with environment, time, and working state. Here we propose a nondestructive method based on acoustoelasticity to measure stresses inside a soft material without invoking the prior knowledge of these constitutive parameters.

The acoustoelastic effect has previously been reported in soft materials, see, e.g., Ref.~\citep{ref9, ref24, ref26}. Interestingly, soft materials can undergo large elastic deformations when subject to mechanical stresses, which alter the shear wave speeds dramatically (\textasciitilde\SI{100}{\percent}) but barely change the speed of the longitudinal wave. That is because it only takes stresses in the \SI{}{kPa} to deform soft solids, and typically, the latter speed ($v_L$, say) is such that $\rho v_{L}^{2}$ (where $\rho$ is the mass density) is in this order of \SI{}{GPa}, while the former speed ($v_T$, say) is such that $\rho v_{T}^{2}$ is in the order of \SI{}{kPa} \citep{ref27}. Technically, the unaffected longitudinal (ultrasound) waves travel \textasciitilde1,000 times faster than shear waves. They provide a unique way to excite (by acoustic beam focusing) and visualize (by ultrasound imaging) shear waves remotely and locally. In this method, we create a supershear moving-load that remotely excites shear waves propagating along two orthogonal directions, and we measure their speeds with a frame rate of \SI{10}{\kilo\hertz}. We validate our method by successfully measuring uniaxial and bending stresses in a hydrogel sample, and tensile stress in a skeletal muscle (which is intrinsically anisotropic due to the preferred direction of the aligned muscle fibers). In these measurements of mechanical stresses, we do not need to know, or use, the constitutive parameters of the materials.

\section*{Results}
\justify

\subsection*{Measuring Mechanical Stresses with Shear Waves}
\justify
Consider a plane shear wave with mechanical displacement $u=u_{0}e^{ik(x_{1}\cos\theta+x_{3}\sin\theta-vt)}$ propagating in an incompressible soft solid subject to in-plane stresses $\sigma_{1}$ and $\sigma_{3}$, where $u_{0}$ is the amplitude that lies in the propagation plane, $v$ is the phase speed, $t$ is the time, $x_{i}(i=1,3)$ is the Cartesian coordinate system aligned with the principal stress, and $k$ is the wave number. The wave vector is $\mathbf{k}=k[\cos\theta,0,\sin\theta]^{T}$ and $\theta$ denotes the angle between $\mathbf{k}$ and the $x_1$ axis. The material can have any form of anisotropy, such as due to initial stress \citep{ref25, ref28, ref29} or fibres reinforcing the solid \citep{ref30}, as long as they are aligned with the principal directions of the stress. In effect, for many tissues, structural anisotropy is co-axial with the stress because collagen fibrils often act to optimise the load bearing capacity \citep{ref31, ref32, ref33}. Inserting the plane wave form into the equations of acoustoelasticity we get (see Supplementary Materials, SM, Notes 1 and 2)
\begin{equation} \label{eq:1}
\rho v^2 = \alpha \cos^2\theta + 2\beta \cos^2\theta \sin^2\theta + \gamma \sin^2\theta,
\end{equation}
where $\alpha=\mathcal{H}_{1313}^{0}$, $2\beta=\mathcal{A}_{1111}^{0}+\mathcal{H}_{3333}^{0}-2\mathcal{H}_{1133}^{0}-2\mathcal{H}_{3113}^{0}$, $\gamma=\mathcal{H}_{3131}^{0}$, and $\mathcal{H}_{piqj}^{0}$ are the components of the Eulerian elastic moduli tensor.

Now consider two shear waves, traveling in two perpendicular directions $\theta=\theta_{0}$ and $\theta=\pi/2+\theta_{0}$ with phase speeds $v_{x}$ and $v_{z}$, respectively, where x and z denote a Cartesian coordinate system aligned with the main axes of the transducer (z is the axial direction and y is the elevational direction). We find that $\rho(v_{x}^{2}-v_{z}^{2})=(\alpha-\gamma)\cos(2\theta_{0})$ according to \eqref{eq:1}, and $\alpha-\gamma=\sigma_{1}-\sigma_{3}$, regardless of the constitutive model and out-of-plane stress (see SM, Notes 1 and 2). Taking the two equations together, we conclude that
\begin{equation} \label{eq:2}
\sigma_{1}-\sigma_{3}=\rho\frac{v_{x}^{2}-v_{z}^{2}}{\cos2\theta_{0}}
\end{equation}
which is the foundation of our method to measure mechanical stresses in soft materials. For the case of uniaxial stress $(\sigma_{3}=0)$, Eq.~\eqref{eq:2} gives direct access to $\sigma_{1}$. While Eq.~\eqref{eq:2} holds for any $\theta_{0}$, we find $\theta_{0}=0$ is the optimized condition for practical measurements. On the one hand, $\theta_{0}=0$ gives the best sensitivity to the stress when the speeds are measured. On the other hand, the group speed $v_{g}\equiv\partial(kv)/\partial k$ is usually accessible in ultrasound shear wave elastography \citep{ref34}, instead of the phase speed $v$ that is involved in Eq.~\eqref{eq:2}. Fortunately, the phase speed and group speed are identical along the principal directions (see SM, Fig.~S1). Therefore, the configuration $\theta_{0}=0$ is still effective in the condition where only the group speed is available. For these reasons, we focus on this orientation in the present study.

\begin{figure}[p] 
    \centering
    \includegraphics[width=0.8\linewidth]{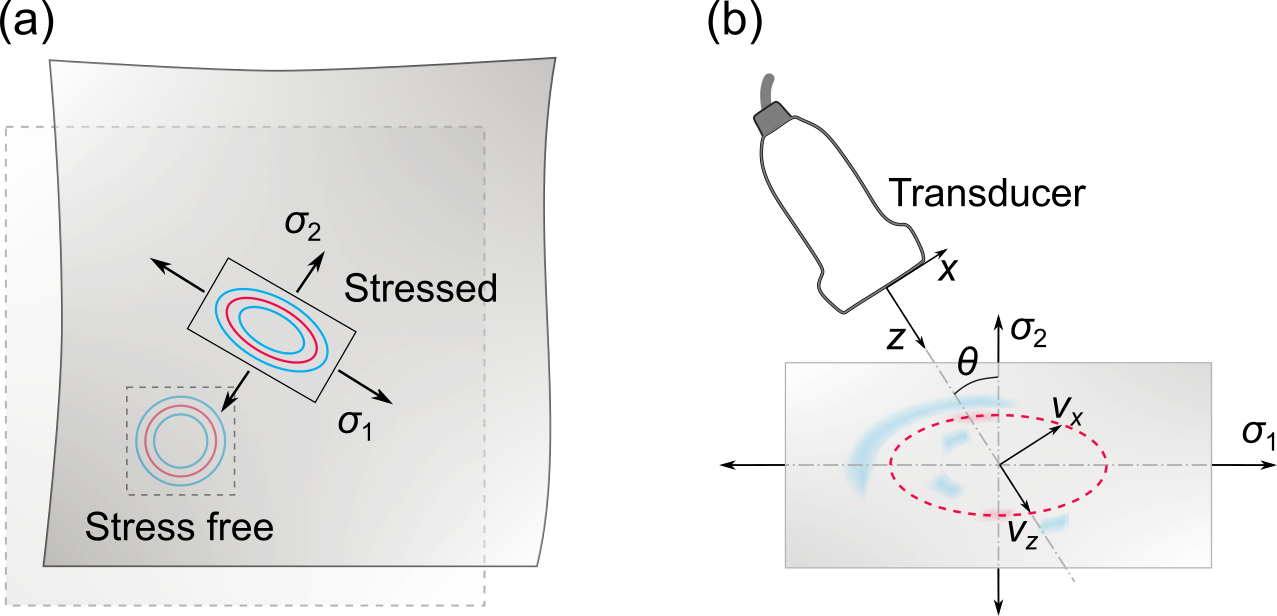}
    \caption{Principle of acoustoelastic imaging. (a) Schematic showing the principal stresses $\sigma_1$ and $\sigma_3$ change the speed of the vertically polarized shear waves, here an isotropic material subject to moderate stress is taken as an example. (b) An ultrasonic transducer with the axial direction ($z$) aligned with the principal direction $x_3$ is used to measure the wave speeds $v_x$ and $v_z$ along the two principal directions. The principal stresses are connected to the two shear wave speeds.}
    \label{fig:1}
\end{figure}

To illustrate, we take an isotropic material as an example. As shown in Fig.~\ref{fig:1}A, the wavefront of the group speed becomes elliptical due to moderate stress (where $2\beta\approx\alpha+\gamma$, see SM, Note 2). The two symmetric axes of the elliptical wavefront are aligned with the principal stresses $\sigma_{1}$ and $\sigma_{3}$, making it easy to identify the directions of the principal stresses from the shape of the wavefront. Figure~\ref{fig:1}B shows the schematic of the ultrasound transducer when $\theta_{0}=0$. Since x and z are aligned with the principal stress direction, the group speed measured along x and z are identical to the phase speeds $v_{x}$ and $v_{z}$, which enables the measurement of mechanical stress according to Eq.~\eqref{eq:2}.

For anisotropic materials, the wave speed is polarization-dependent. In our theory, $u_{0}$ lies in the wave propagation plane, indicating our method relies on the in-plane polarized shear waves (i.e., vertically polarized shear waves, SV). A representative plot for the phase and group speeds of SV shear waves in an anisotropic material subject to uniaxial stress can be found in Fig.~S1B.

\subsection*{Generating Shear Waves propagating in perpendicular directions with programmed acoustic radiation force}
\begin{figure}[p]
    \centering
    \includegraphics[width=\linewidth]{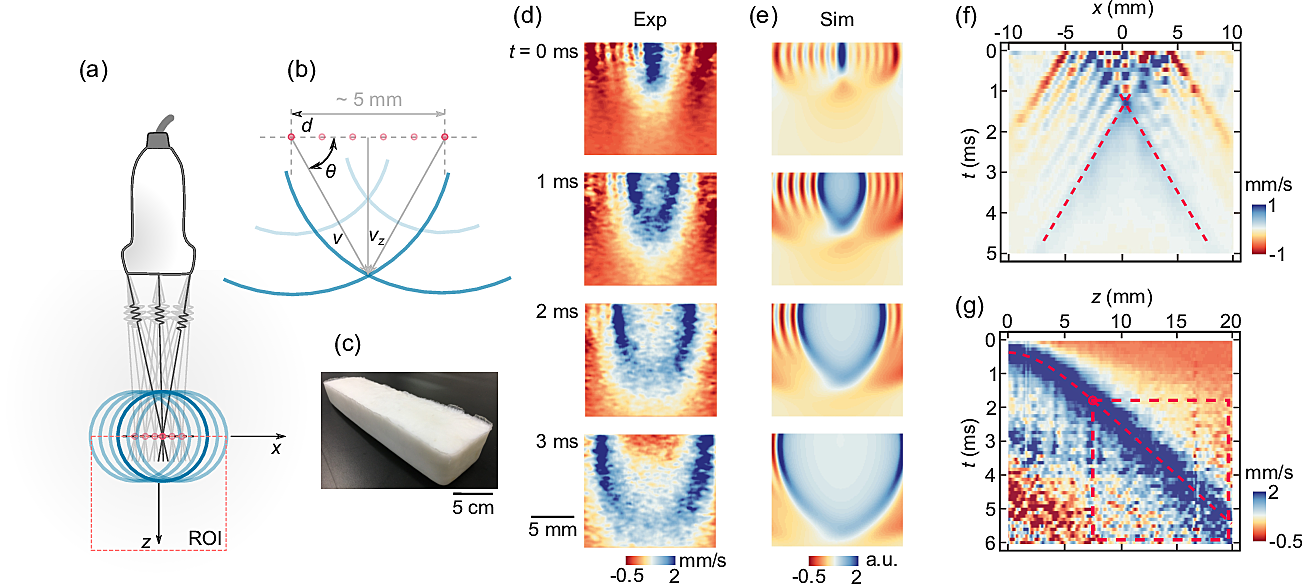}
    \caption{Acoustoelastic imaging using ultrasound shear wave elastography. (a) Schematic of the experimental setup. An ultrasound beam focuses successively from left to right along the $x$ axis at six locations inside the material separated by distance $d = \SI{1}{\milli\meter}$ to excite multiple shear waves. Interference of the shear waves gives rise to a strong vertical shear wave (along the $z$ axis). Wave propagation in the region-of-interest (ROI) is measured by plane wave ultrasound imaging. (b) Schematic showing the propagation of the interference at $(2.5d, z)$, with speed $(z/\sqrt{z^2 + (2.5d)^2})v$. (c) Photograph of the hydrogel sample at rest. (d) Snapshots showing the shear wave propagation in the ROI. The maps depict the vertical particle velocity fields. (e) Finite element simulations of the shear wave propagation. (f) and (G) Spatiotemporal maps of the lateral (along $x$) and vertical (along $z$) shear waves. (g) shows that the shear wave speed is constant only when the shear wave propagates far away ($z > \SI{7}{\milli\meter}$, the dashed square), in line with the theoretical prediction $(z/\sqrt{z^2 + (2.5d)^2})v \rightarrow v_z$ for large $z$. The shear wave speeds $v_x$ and $v_z$ are measured from (f) and (g), respectively, by the Radon transformations (see Fig.~S4).}
    \label{fig:2}
\end{figure}

\justify
Our experimental setup to generate two shear waves propagating perpendicularly to each other, shown in Fig.~\ref{fig:2}A, was based on a medical ultrasound imaging system (see Methods). The ultrasound transducer sent \SI{7}{\mega\hertz} ultrasound waves that were used to excite and detect shear waves in soft materials. In effect, the absorption of the ultrasound waves leads to a transfer of momentum to the soft materials, giving rise to the acoustic radiation force (ARF). A focused ultrasound beam can deliver the ARF locally, resulting in a Gaussian-shaped body force at the focus (see SM, Fig.~S2B). Micrometer amplitude shear waves traveling perpendicular to the ultrasound beam (x axis) are then generated by the ARF, and measuring their speed enables what is called shear wave elastography \citep{ref35, ref36}. However, with a standard setup, shear waves traveling along the beam direction (z axis) are not easily detectable because they are small and attenuate rapidly \citep{ref37} (see supplementary movies S1B and S2B for simulation and experimental results, respectively).

To excite the lateral and vertical shear waves simultaneously, we present a new programming method that successively focuses the ultrasound beam at six locations (duration at each location \textasciitilde\SI{43}{\micro\second}), separated by a distance of $d=\SI{1}{\milli\meter}$ along the lateral direction x, as shown in Fig.~\ref{fig:2}A. These ARFs mimic a laterally moving load with a supershear wave speed (the ratio of the moving speed and the shear wave speed, i.e., the Mach number, is 10). The shear waves generated by the moving load mutually interfere following the Huygens-Fresnel principle, which significantly enlarges the amplitude of the vertical wave. The vertical shear waves are primarily vertically polarized, i.e., the so-call longitudinal shear waves, which have been utilized in ultrasound elastography of the livers \citep{ref38, ref39}. Approximately \SI{0.3}{\milli\second} after the wave excitation, unfocused ultrasound beams are sent by the same ultrasound transducer to perform ultrafast ultrasound imaging \citep{ref40}, which records the shear wave propagation in the region-of-interest (ROI) at a rate of 10,000 frames per second.

We tested our experimental setup on a polyvinyl alcohol (PVA) hydrogel (mass density $\rho\sim\SI{1}{\gram\per\centi\meter\cubed}$, initial shear modulus \textasciitilde \SI{8.6}{\kilo Pa}, see Methods). The approximate size is $29 \times \SI{6}{\centi\meter\squared}$ cross-section and \SI{4}{\centi\meter} depth (Fig.~\ref{fig:2}C). Figure~\ref{fig:2}D depicts the snapshots of the shear wave propagation in the sample, and clearly shows that the lateral and vertical shear waves are generated simultaneously, in excellent agreement with the finite element simulations (see Methods) shown in Fig.~\ref{fig:2}E and supplementary movie S1A. As for anisotropic materials, we also performed three-dimensional finite element simulations to confirm that the SV shear waves are primarily excited using the programmed ARFs, and the lateral and vertical SV shear waves are generated simultaneously (see Fig.~S3).

To measure the shear wave speeds, we extract the spatiotemporal data along the lateral (x axis) and vertical (z axis) directions, respectively. As shown in Fig.~\ref{fig:2}F, six shear waves propagate to the left and to the right, with a linear wavefront which suggests that the wave speed $v_{x}$ is constant. However, the vertical shear waves gradually decelerate from the near-field to the far-field (Fig.~\ref{fig:2}G), with the measured speed $v_{z}$ approximately following $v \frac{z}{\sqrt{z^{2}+(2.5d)^{2}}}$, where $v$ is the shear wave speed along $\theta=\tan^{-1}(\frac{z}{2.5d})$. This is expected and is likely due to the wave interference pattern depicted in Fig.~\ref{fig:2}B. Note that for large enough z we have $v \frac{z}{\sqrt{z^{2}+(2.5d)^{2}}} \sim v$ and $v$ should be the speed of the vertical shear wave we want to measure. For this reason we only use the data for $z > \SI{7}{\milli\meter}$ (the dashed square in Fig.~\ref{fig:2}G) in the subsequent analysis.

To derive the group velocities in a robust way, we apply the Radon transformation \citep{ref41} to the spatiotemporal data shown in Figs.~\ref{fig:2}F-G to compute $v_{x}$ and $v_{z}$ (for the lateral direction x, a directional filter is performed to the spatiotemporal data before the Radon transformation, see SM, Note 3 and Fig.~S4). In the absence of mechanical stress, we get $v_{x}=\SI{2.81 \pm 0.05}{\meter\per\second}$ and $v_{z}=\SI{2.82 \pm 0.06}{\meter\per\second}$, which agrees with the theoretical prediction that $v_{x}=v_{z}$ in the absence of mechanical stress. The initial shear modulus derived from the shear wave speeds is $\mu=\SI{8.46 \pm 0.33}{\kilo Pa}$, in agreement with the mechanical characterization performed by indentation tests (shear modulus $\SI{8.6 \pm 0.3}{\kilo Pa}$, see SM, Note 4).

\subsection*{Measuring Stresses in Hydrogel and Muscle without the Knowledge of their constitutive parameters}
\begin{figure}[p]
    \centering
    \includegraphics[width=\linewidth]{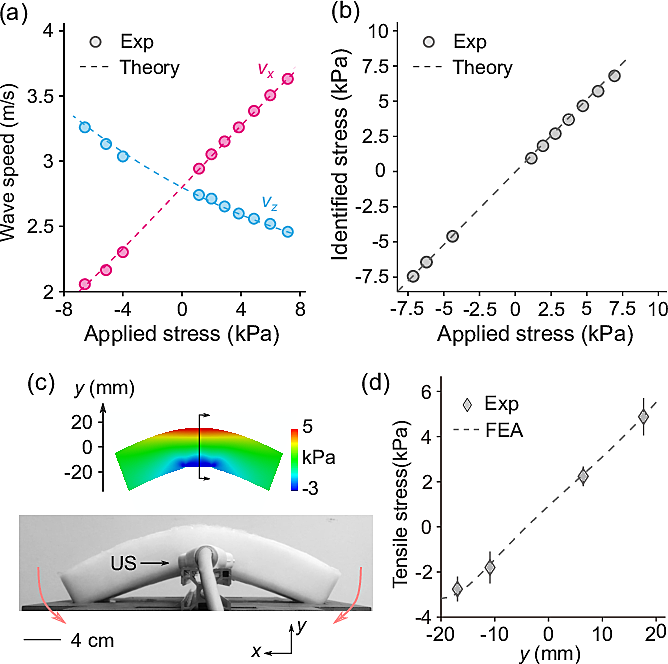}
    \caption{Acoustoelastic imaging of a soft material. (a) Shear wave speeds measured in a hydrogel subject to a uniaxial stress. (b) Comparison of identified stress with the applied stress. Dashed line: $45^\circ$ line for visual guide. (c) Photograph showing the sample under bending deformation and Finite Element computation of the bending stress. (d) Bending stress measured by acoustoelastic imaging and comparison with theory. Error bars denote the standard deviations of five measurements.}
    \label{fig:3}
\end{figure}

\justify
For our first test to demonstrate the usefulness of our theory and method, we applied a uniaxial stress to the hydrogel sample $\sigma_{1}$ along the x direction and then measured $v_{x}$ and $v_{z}$. As shown in Fig.~\ref{fig:3}A, the tensile/compressive stress increases/decreases $v_{x}$ but decreases/increases $v_{z}$. The identified stress shows a good agreement with the applied stress, with maximum error \textasciitilde \SI{5}{\percent} (Fig.~\ref{fig:3}B). Further, we measured the stress induced by a bending deformation of the hydrogel sample. As shown in Fig.~\ref{fig:3}C, we applied a \SI{4}{\centi\meter} deflection to bend the sample, which resulted in an approximately linear stress field across the thickness of the sample (see the simulation in Fig.~\ref{fig:3}C). We perform measurements within four planes parallel to the neutral plane of zero stress, at y = -20, -14.7, 12.8, \SI{20}{\milli\meter}. Figure~\ref{fig:3}D shows the stresses measured at different locations, which agree with the theoretical values obtained using finite element simulations.

\begin{figure}[p]
    \centering
    \includegraphics[width=\linewidth]{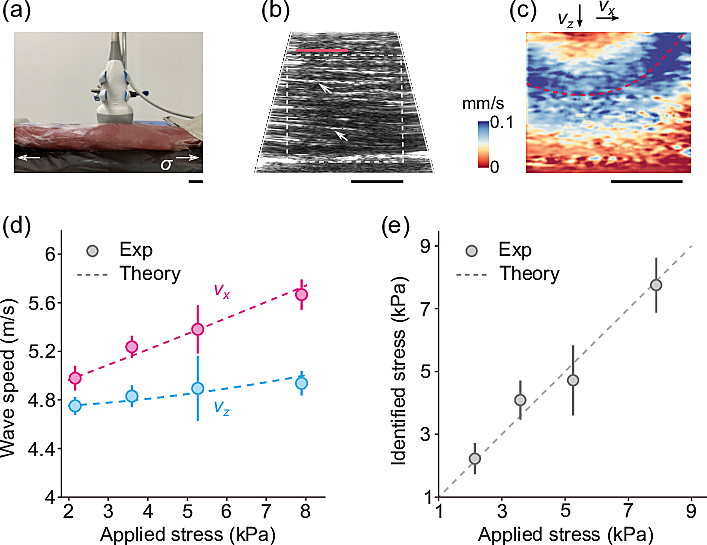}
    \caption{Acoustoelastic imaging of a skeletal muscle. Scale bar, \SI{1}{\centi\meter}. (a) Photograph of the skeletal muscle. (b) Grayscale B-mode image of the sample. In this view the muscle fibers (some indicated by the arrows) and the applied stress are along the horizontal direction. The Acoustic Radiation Forces are applied along the red line. Dashed square, ROI where the wave speeds are measured. (c) A representative snapshot (\textasciitilde\SI{2.6}{\milli\second} after ARFs push) of the wave propagation when the applied stress is \textasciitilde\SI{3.6}{\kilo Pa}. (d) Shear wave speeds measured at different levels of stress. Markers, experiment. Error bar, the standard deviations of five measurements. Dashed lines, theoretical curves that are obtained using a phenomenological model (see SM, Note 5). (e) Comparison between applied stress and identified stress. Dashed line, the $45^\circ$ line for visual guide.}
    \label{fig:4}
\end{figure}

We proceed to demonstrate the effectiveness of our method in probing the mechanical stresses in anisotropic soft tissues. To this end, we performed ex vivo measurements on a sample of porcine skeletal muscle, as shown in Fig.~\ref{fig:4}A. The elastic deformation of the skeletal muscle can be captured using a transversely isotropic model reflecting the preferential orientation of the muscle fibers, as shown by the ultrasound B-mode image (Fig.~\ref{fig:4}B). In this experiment we applied a tensile stress along the muscle fibers using several weights (each weight \textasciitilde\SI{500}{\gram}), mimicking a passive stretch of the skeletal muscle \citep{ref42}. Figure~\ref{fig:4}C shows a representative snapshot (\textasciitilde\SI{2.6}{\milli\second} after the AFRs push) of the shear wave propagation, when the applied stress is \textasciitilde \SI{3.6}{\kilo Pa}. The ARFs are applied on the left side of the ROI, and then $v_{x}$ is measured for the shear wave propagating from left to right. Compared with the hydrogel, it is apparent that the wavefronts are broader, because of a larger shear wave speed, and that there is a stronger dissipation (see SM, Note 5 for mechanical characterization of the skeletal muscle).

Figure~\ref{fig:4}D shows the velocities $v_{x}$ and $v_{z}$ obtained when the muscle is subject to different levels of mechanical stresses. The measurement uncertainties on the wave speeds are larger compared with the measurements on the hydrogel sample, due to the broader wavefronts. As expected intuitively, the wave speed $v_{x}$ along the tension/fiber direction increases with the tensile stress. Notably, the shear wave speed $v_{z}$ in the skeletal muscle increases with tension along x, in contrast to the isotropic hydrogel where $v_{z}$ decreases. This is likely due to the nonlinear elastic response of the skeletal muscle, which makes it stiffer when increasing the tension \citep{ref43, ref44}. In the analysis, we find a phenomenological model incorporating exponentially stiffening effects (see SM, Note 5) fits the experimental data, as shown in Fig.~\ref{fig:4}D. The nontrivial acoustoelastic properties of the muscle again highlight the key advantage of our acoustoelastic imaging method: no acoustoelastic parameters of the materials were needed to predict the stress. We simply derive the tensile stresses from the shear wave speeds, as shown in Fig.~\ref{fig:4}E. The stress identified by our method shows a good agreement (maximum error \textasciitilde\SI{15}{\percent}) with the applied stress. We attribute the larger error to the viscoelasticity of the biological sample.

\section*{Discussion and Conclusion}
\justify
Based on the acoustoelastic principle we proposed a theory and a method to probe mechanical stresses in soft materials without the prior knowledge of their constitutive parameters, in contrast to the existing methods presented to date. A key step to realise our method was to program multiple ARFs to mimic a supershear moving-load generating shear waves in two mutually perpendicular directions. We were then able to obtain the speeds of both waves by ultrasonic imaging, which, according to our theory, allowed us to measure the mechanical stresses remotely. Hence we successfully measured the spatial variation of a bending stress in a hydrogel and of a tensile stress in a passively stretched muscle, which is intrinsically anisotropic. The stretched muscle test illustrates how our method works even in the presence of structural anisotropy when it is aligned with the stress.

The effect of the viscoelasticity of soft materials on the proposed method deserves a careful discussion. As indicated by our experiments on skeletal muscle, inaccuracies may appear when neglecting viscosity. For high enough frequencies, biological
tissues exhibit frequency-dependent response due to viscosity, which in turn may affect the predictions of our method. To address this issue, we invoke the quasi-linear viscoelasticity theory (QLV), which models the stress relaxation with a Prony series, $\mu(t)=\mu_{0}[1-\sum_{i=1}^{n}g_{i}(1-e^{-t/\tau_{i}})]$, where $\mu(t)$ is the relaxation shear modulus in response to a step constant strain, $\mu_{0}$ is the instantaneous shear modulus, $\tau_{i}$ is a characteristic relaxation time, and $g_{i}$ is a dimensionless relaxation modulus $(i=1,2,...,n)$. For simplicity we take $n=1$ and find that this model fits well the viscoelastic dispersion of shear waves in skeletal muscle over the \SIrange{100}{500}{\hertz} range, with $g_{1}=0.79$ and $\tau_{1}=\SI{0.49}{\milli\second}$ (see Fig.~S6E). We then use this model to evaluate the effect of viscoelasticity on the identified mechanical stresses based on a recently proposed acousto-visco-elastic theory \citep{ref45}. The results show that over a broad frequency range (\SIrange{10}{1000}{\hertz}) the stress is underestimated when viscoelasticity comes into play (see SM, Note 5 and Fig.~S7). However, in our method we use the group velocity of the shear waves (\SI{4}{dB} bandwidth from \SIrange{100}{1000}{\hertz}, see Fig.~S8), and the average error over the frequency band is \textasciitilde\SI{16}{\percent}, consistent with our measurements. For soft materials where the extent of stress relaxation is less than \textasciitilde\SI{50}{\percent}, which covers a wide range of soft materials including most hydrogels and soft tissues, our analysis indicates that shear wave dispersion caused by viscosity has negligible effect on mechanical stresses measured with the reported acoustoelastic imaging method (maximum error is less than \SI{10}{\percent}).

Measuring the constitutive parameters of a soft material in situ is indeed challenging because the parameters change with time, environment, and from one working state to another. By bypassing this difficulty, our constitutive parameter-free theory and method to probe mechanical stresses in a non-destructive manner should find broad applications across different disciplines including, but not limited to, biomedical engineering, biology, medicine, materials science and soft matter physics.

\section*{Materials and Methods}
\justify

\subsection*{Ultrasound Setup}
\justify
Our ultrasound experimental system was built on the Vantage 64LE system (Verasonics Inc., Kirkland WA, USA). The central frequency, pitch, and element number of the ultrasound transducer (L9-4, JiaRui Electronics Technology Co., Shenzhen, China) used in our experiments were \SI{7}{\mega\hertz}, \SI{0.3}{\milli\meter}, and 128, respectively. The imaging sequence of the ultrasound experiment is depicted in Fig.~S2A. In the excitation stage, the focused ultrasound beams were generated by 32 elements (voltage \textasciitilde \SI{10}{V}, aperture size \textasciitilde \SI{10}{\milli\meter}, and uniform apodization). The focus was \textasciitilde \SI{13}{\milli\meter} away from the transducer. In the imaging stage, while all the 128 elements (voltage \textasciitilde \SI{10}{V}, aperture size \textasciitilde \SI{40}{\milli\meter}, and uniform apodization) were used to transmit unfocused ultrasound beams, only the 64 elements at the center of the transducer were used as receivers. The ultrasound in-phase and quadrature (IQ) signals during the wave propagation were acquired at a frame rate of \SI{10}{\kilo\hertz}. The plane wave imaging with delay and sum beamforming was adopted to reconstruct each frame \citep{ref46}. The particle velocity field was calculated offline based on the Loupas’ estimator \citep{ref47} using a kernel size of 5$\times$2 (\SI{0.275}{\milli\meter} in $x$ and \SI{0.2}{\milli\second} in $t$). A spatial filter (mean filter) with a kernel size of 8$\times$8 (\SI{0.87}{\milli\meter} in $x$ and \SI{0.44}{\milli\meter} in $z$) was then employed to reduce the noise of the particle velocity. For all the experiments, ten successive measurements (\textasciitilde \SI{56}{\milli\second}) were performed and the average of the measurements was taken to improve the signal-to-noise ratio.

\subsection*{Hydrogel Phantom Preparation}
\justify
The hydrogel consisted of \SI{10}{\percent} polyvinyl alcohol (PVA), \SI{3}{\percent} cellulose and \SI{87}{\percent} deionized water by weights. We dissolved the PVA powder (sigma Aldrich 341584, Shanghai, China) into \SI{80}{\celsius} water. We then added cellulose powder (Sigma-Aldrich S3504, Shanghai, China) into the solution and fully stirred the solution to get a suspension of the cellulose powder. The cellulose particles act as ultrasonic scatterers to enhance the imaging contrast. We poured the suspension into a square plastic box (length \textasciitilde \SI{30}{\centi\meter}, width \textasciitilde \SI{7}{\centi\meter}, and height \textasciitilde \SI{4}{\centi\meter}), and then cooled the suspension to room temperature (\textasciitilde \SI{20}{\celsius}) before putting it into a \SI{-20}{\celsius} freezer. We froze the sample for 12 hours and then thawed it at room temperature for another 12 hours. The stiffness of the sample can be tuned by freezing/thawing (F/W) cycles \citep{ref48}. The hydrogel sample used in this study underwent two F/W cycles. We performed indentation tests on the hydrogel and measured the dispersion relation of the Rayleigh surface waves to characterize its elastic and viscoelastic properties (see SM, Note 4 and Fig.~S5).

\subsection*{Finite Element Analysis}
\justify
The Finite Element analyses (FEA) were performed using Abaqus (Abaqus 6.14, Dassault Systèmes®). For the shear wave generation in isotropic materials, we used a plane strain model with an in-plane size of $50 \times \SI{50}{\milli\meter\squared}$. The acoustic radiation force was modeled as a body force with a Gaussian shape of the form
\begin{equation} \label{eq:3}
\mathbf{f} = \mathbf{f}_0 \exp \left( -\frac{(x - x^{(i)})^2}{2r_x^2} - \frac{(z - z^{(i)})^2}{2r_z^2} \right),
\end{equation}
where $\mathbf{f}_0$ is the magnitude of the force, with direction parallel to the ultrasound beam and magnitude small enough to generate small-amplitude waves, and $(x^{(i)}, z^{(i)})$ ($i = 1, 2, \dots , 6$) are the coordinates of the six focal points. Also, the parameters $r_x = \SI{0.5}{\milli\meter}$ and $r_z = \SI{1.0}{\milli\meter}$ were taken according to the experimental data shown in Figs~S2B and C. We used a uniform mesh grid (element size \SI{0.1}{\milli\meter}) and the CPE8RH element (plane strain, 8-node biquadratic, reduced integration, hybrid with linear pressure). Other parameters used in the simulations and the post analyses were consistent with our experimental setup.

For the shear wave generated in anisotropic materials by programmed ARFs, we built a three dimensional model with Abaqus/explicit. We used similar geometry parameters as the plane model for isotropic materials, but extended the model thickness to be \SI{30}{\milli\meter} along the elevational direction ($y$ axis). The Gaussian radius of the acoustic radiation force in $y$ axis is $r_y = r_x$. We used the C3D8 (8-node linear brick, hybrid with constant pressure) element in the simulation and the average mesh size for the 3D model is about $0.1 \times 0.1 \times \SI{0.1}{\milli\meter}$.

In the FEA of the bending stress, we built a plane-stress model that was \SI{30}{\centi\meter} long and \SI{4}{\centi\meter} wide. The size of the model was consistent with our physical sample. We fixed the lower left and right corners of the sample and prescribed the displacement (\SI{6}{\centi\meter}) at the middle of the lower boundary. We used a uniform mesh (\SI{0.5}{\centi\meter}) and the CPS8R element (plane stress, 8-node biquadratic, reduced integration).

\section*{Acknowledgments}
\justify
We gratefully acknowledge support from the National Natural Science Foundation of China (Grants Nos. 11972206 and 11921002). ALG is grateful for support from the European Commission - Horizon 2020 / H2020 - Shift2Rail for A.L.G.

\clearpage
\appendix
\section*{Supplementary Materials}
\justify
The PDF file includes:
Supplementary Text
Figs. S1 to S7
Legends for Movies S1 to S2

Other Supplementary Material for this manuscript includes the following:
Movies S1 to S2

\subsection*{Supplementary Note 1: The stress identity}
\justify
Here we prove the following identity, used in the paper to connect wave speeds with stress:
\begin{equation} \label{eq:S1}
\sigma_{11} - \sigma_{33} = \mathcal{A}^0_{1313} - \mathcal{A}^0_{3131}.
\end{equation}
In the paper we write these moduli as $\alpha = \mathcal{A}^0_{1313}$ and $\gamma = \mathcal{A}^0_{3131}$, and considered scenarios where the components of the Cauchy stress $\sigma_{11}$ and $\sigma_{33}$ are the principal stresses $\sigma_{1}$ and $\sigma_{3}$, respectively. Here $\mathcal{A}^0_{piqj}$ are the Cartesian components of the Eulerian elasticity tensor. For incompressible solids, they are determined from the strain energy function $W$ and the deformation gradient tensor with components $F_{iJ}$ as \citep{ref28, ref49}
\begin{equation} \label{eq:S2}
\mathcal{A}^0_{piqj} = (\sigma_{pq} + \bar{p}\delta_{pq})\delta_{ij} + 4F_{pP}F_{qQ} \frac{\partial^2 W}{\partial C_{IP}\partial C_{QJ}} F_{iI}F_{jJ},
\end{equation}
where $C_{IJ} = F_{kI}F_{kJ}$, summation over repeated indices is implied, and $\delta_{ij}$ is the Kronecker delta. Hence
\begin{align}
\mathcal{A}^0_{1313} &= \sigma_{11} + \bar{p} + 4F_{1P}F_{1Q} \frac{\partial^2 W}{\partial C_{IP}\partial C_{QJ}} F_{3I}F_{3J}, \label{eq:S3} \\
\mathcal{A}^0_{3131} &= \sigma_{33} + \bar{p} + 4F_{3P}F_{3Q} \frac{\partial^2 W}{\partial C_{IP}\partial C_{QJ}} F_{1I}F_{1J} \nonumber \\
                 &= \sigma_{33} + \bar{p} + 4F_{1P}F_{1Q} \frac{\partial^2 W}{\partial C_{PI}\partial C_{JQ}} F_{3I}F_{3J}, \label{eq:S4}
\end{align}
where for the last equation we swapped the dummy variables $I \leftrightarrow P$ and $Q \leftrightarrow J$, and then we used the symmetries $C_{IP} = C_{PI}$. By subtraction we obtain the identity \eqref{eq:S1}.

Often the stress is modeled as being caused by a finite elastic deformation from a stress-free configuration. When instead, we consider small elastic waves in an initially stressed reference where the initial stress, denoted by $\tau_{ij}$ is due to any origin, then in the above we would take $F_{pP} = \delta_{pP}$ \citep{ref28, ref29}, and the identity would still hold. For future reference, we recall that the Cauchy stress is computed as \citep{ref49}
\begin{equation} \label{eq:S5}
\sigma_{ij} = F_{iK} \frac{\partial W}{\partial F_{jK}} - \bar{p}\delta_{ij},
\end{equation}
where $\bar{p}$ is a Lagrange multiplier due to the constraint of incompressibility.

\subsection*{Supplementary Note 2: The equation of wave motion}
\justify
Here we relate the wave speeds to the moduli appearing in the stress identity \eqref{eq:S1}. We start with the equation of motion for plane shear waves of the form $\mathbf{u} = \mathbf{u}_0 e^{ik(\mathbf{n}\cdot\mathbf{x}-vt)}$, which is given by Equation (5.16) in Ref.~\citep{ref50}:
\begin{equation} \label{eq:S6}
(\mathbf{I} - \mathbf{n}\mathbf{n}^T)\mathbf{Q}(\mathbf{n})(\mathbf{I} - \mathbf{n}\mathbf{n}^T)\mathbf{u}_0 = \rho v^2 \mathbf{u}_0,
\end{equation}
where $\mathbf{x} = (x_1, x_2, x_3)$, $\mathbf{n} = (n_1, n_2, n_3)$, $Q_{ij}(\mathbf{n}) = \mathcal{A}^0_{piqj}n_p n_q$, and $\mathbf{u}_0$ is a unit vector along the direction of polarization (orthogonal to $\mathbf{n}$, the unit vector along the direction of propagation). Then its wave speed $v$ is given by
\begin{equation} \label{eq:S7}
\rho v^2 = \mathbf{u}_0^T \mathbf{Q}(\mathbf{n})\mathbf{u}_0.
\end{equation}
Let $v_x$ and $v_z$ be the speeds of the shear waves when $\mathbf{n} = (1, 0, 0)$, $\mathbf{u}_0 = (0, 0, 1)$, and $\mathbf{n} = (0, 0, 1)$, $\mathbf{u}_0 = (1, 0, 0)$, respectively. From the above it follows that
\begin{equation} \label{eq:S8}
\rho v_x^2 = \mathcal{A}^0_{1313}, \quad \rho v_z^2 = \mathcal{A}^0_{3131}.
\end{equation}
To guarantee that there are two shear waves with speeds \eqref{eq:S8}, that satisfy the equation of motion \eqref{eq:S6}, we assume that all forms of anisotropy are coaxial with the deformation tensor $\mathbf{C} = \mathbf{F}\mathbf{F}^T$. Different types of anisotropy, such as the ones captured by an initial stress tensor $\bm{\tau}$ \citep{ref28, ref29, ref30} or a structural anisotropy tensor $\mathbf{M}\mathbf{M}^T$ (where $\mathbf{M}$ is a unit vector along the preferred direction in the reference configuration for transversely isotropic materials, see for example Ref.~\citep{ref50}), can be included in the strain-energy $W$, from which we can deduce the moduli $\mathcal{A}^0_{piqj}$ with \eqref{eq:S2}. For example, $\bm{\tau}$ and $\mathbf{M}\mathbf{M}^T$ are coaxial with $\mathbf{C}$, and themselves, when $\mathbf{C}\bm{\tau} = \bm{\tau}\mathbf{C}$, $\mathbf{C}\mathbf{M}\mathbf{M}^T = \mathbf{M}\mathbf{M}^T\mathbf{C}$, and $\bm{\tau}\mathbf{M}\mathbf{M}^T = \mathbf{M}\mathbf{M}^T\bm{\tau}$. This condition implies, for example, that $\mathbf{M}$ is aligned with the principal directions of the initial stress $\bm{\tau}$ and the final stress $\bm{\sigma}$.

In more detail, $W$ can be written as a sum and multiplication of terms of the form $\text{tr}(\mathbf{A}\mathbf{C}^n\mathbf{B})$ for integer $n$ where $\mathbf{A}$ and $\mathbf{B}$ are some multiplication of anisotropy tensors such as $\bm{\tau}$ and $\mathbf{M}\mathbf{M}^T$. When all these tensors are coaxial, and we choose a coordinate system align with their axes, we find that
\begin{equation} \label{eq:S9}
\mathcal{A}^0_{piqj} = 0 \quad \text{unless} \quad
\begin{cases}
p = i \quad \& \quad q = j, \text{ or} \\
p = q \quad \& \quad i = j, \text{ or} \\
p = j \quad \& \quad q = i.
\end{cases}
\end{equation}
By assuming the above, we can deduce which elastic shear waves can give us access to the stress identity \eqref{eq:S1}. Let $\mathbf{n} = (\cos\theta, \sin\theta, 0)$ and $\mathbf{u}_0 = (-\sin\theta, \cos\theta, 0)$, which substituted into \eqref{eq:S7} leads to
\begin{equation} \label{eq:S10}
\rho v^2 = \alpha \cos^4\theta + 2\beta \cos^2\theta \sin^2\theta + \gamma \sin^4\theta,
\end{equation}
where the instantaneous moduli $\alpha, \beta, \gamma$ are defined as $\alpha = \mathcal{A}^0_{1313}$, $2\beta = \mathcal{A}^0_{1111} + \mathcal{A}^0_{3333} - 2\mathcal{A}^0_{1133} - 2\mathcal{A}^0_{3113}$, $\gamma = \mathcal{A}^0_{3131}$. Note this is the same result as deduced in \citep{ref25, ref49, ref51} with the difference that here we showed that it holds in general when \eqref{eq:S9} holds. This justifies how and when our method applies to anisotropic solids under stress.

Now consider two shear waves, one with propagation direction $\theta = \theta_0$ and another with $\theta = \pm\pi/2 \pm \theta_0$ with the speeds $v_x$ and $v_z$, respectively. Then, according to Eq.~\eqref{eq:S10} and \eqref{eq:S1}, we find that
\begin{equation} \label{eq:S11}
\sigma_1 - \sigma_2 = \rho \frac{v_x^2 - v_z^2}{\cos 2\theta_0},
\end{equation}
a generalization of the result established in \citep{ref25} for isotropic solids.

As we discussed in the main text, the group velocities $\mathbf{v}_g$ are available in many shear wave elastography experiments, instead of the phase speed given in Eq.~\eqref{eq:S10}. According to the definition $\mathbf{v}_g = \partial(kv)/\partial\mathbf{k}$, where $\mathbf{k} = k\mathbf{n}$ denotes the wave vector, the group velocity can be obtained from Eq.~\eqref{eq:S10}. It can be shown the phase and group speed are identical in the principal directions ($\theta = 0$). Interestingly, for isotropic materials subject to moderate stress, we have $2\beta \approx \alpha + \gamma$ \citep{ref51}, resulting in
\begin{equation} \label{eq:S12}
\rho v^2 = \alpha \cos^2\theta + \gamma \sin^2\theta.
\end{equation}
From Eq.~\eqref{eq:S12}, we get $v_{g1} = \frac{\alpha \cos\theta}{\rho v}$ and $v_{g3} = \frac{\gamma \sin\theta}{\rho v}$, and thus
\begin{equation} \label{eq:S13}
\frac{v_{g1}^2}{\alpha/\rho} + \frac{v_{g3}^2}{\gamma/\rho} = 1,
\end{equation}
an elliptical wavefront. This elliptical wavefront has also been revealed by Rouze et al.~\citep{ref34} (see the case of Mooney-Rivlin material, where $2\beta = \alpha+\gamma$ always holds regardless of the stress level). However, for other constitutive models such as the Arruda–Boyce model, Rouze et al.~\citep{ref34} show that cusp structures in wavefront may emerge in isotropic materials when sufficiently large stress is applied. These cusps are usually induced by structural anisotropy of materials, as shown in Fig.~S1.

\subsection*{Supplementary Note 3: Measurement of the lateral shear wave speed}
\justify
We performed two-dimensional Fourier transforms on Fig.~S4A to get the frequency-wavenumber domain data, as shown in Fig.~S4B. To identify the left-to-right (LR) shear waves, we performed an inverse Fourier transform to the data in the first and third quadrants (and set the data points in the second and fourth quadrants to zero), as shown in Fig.~S4C. Similarly, the right-to-left (RL) shear waves were obtained by inverse Fourier transform on the data in the second and fourth quadrants (Fig.~S4D). We then performed Radon transformations to the spatiotemporal data to obtain the shear wave group velocity. The Radon transform sums the intensity of pixels in a spatiotemporal map along projections with different slopes (denoted by $\tan\Theta$) and intercepts. The optimal projection is identified by the peak Radon sum \citep{ref41}. For the lateral shear waves, the six wavefronts induced by the six ARF pushes are parallel, resulting in multiple peaks in the Radon sum (Figs.~S4E and F). Therefore, we summed the absolute values of the Radon sums obtained from the projections with the same slopes (each column of the Radon sums), as shown in Figs.~S4G and H. We identified the maxima in Figs.~S4G and H, respectively, to get the group velocities of the LR and RL shear waves, i.e., $\sim |\tan 67^\circ| \frac{\Delta x}{\Delta t}$ and $\sim |\tan 113^\circ| \frac{\Delta x}{\Delta t}$, respectively, where $\Delta x = \SI{0.1}{\milli\meter}$ and $\Delta t = \SI{0.1}{\milli\second}$ are the grid size of spatiotemporal maps. Finally we reported the average of the two optical group velocities as the value of $v_x$.

\subsection*{Supplementary Note 4: Hydrogel sample characterization}
\justify
The hydrogel consists of \SI{10}{\percent} polyvinyl alcohol (PVA), \SI{3}{\percent} cellulose and \SI{87}{\percent} deionized water by weights. We dissolved the PVA powder (sigma Aldrich 341584, Shanghai, China) into \SI{80}{\celsius} water. We then added cellulose powder (Sigma-Aldrich S3504, Shanghai, China) into the solution and fully stirred the solution to get a suspension of the cellulose powder. The cellulose particles act as ultrasonic scatterers to enhance the imaging contrast. We poured the suspension into a square plastic box (length \textasciitilde \SI{30}{\centi\meter}, width \textasciitilde \SI{7}{\centi\meter}, and height \textasciitilde \SI{4}{\centi\meter}), and then cooled the suspension to room temperature (\textasciitilde \SI{20}{\celsius}) before putting it into a \SI{-20}{\celsius} freezer. We froze the sample for 12 hours and then thawed it at room temperature for another 12 hours. The stiffness of the sample can be tuned by the freezing/thawing (F/W) cycles \citep{ref48}. The hydrogel sample used in this study underwent two F/W cycles.

We performed indentation tests (Fig.~S5A) to characterize the viscoelastic properties of the hydrogel sample. To get the long-term modulus, we performed three indentation tests using a low loading rate (\textasciitilde \SI{0.1}{\milli\meter\per\second}), as shown in Fig.~S5B. The long-term shear modulus $\mu_\infty = \mu(t \rightarrow +\infty)$ can be obtained by fitting the loading curve with the formula
\begin{equation} \label{eq:S14}
F = \frac{16}{9} \mu_\infty R^{1/2} h^{3/2},
\end{equation}
where $R \approx \SI{7.5}{\milli\meter}$ is the radius of the indenter, $F$ is the force, and $h$ is the indentation depth. As shown in Fig.~S5B, the best fitting gives $\mu_\infty = \SI{8.6 \pm 0.3}{\kilo Pa}$. We then increased the loading rate (\textasciitilde \SI{100}{\milli\meter\per\second}) and measured the stress relaxation when holding the indentation depth at \textasciitilde \SI{5}{\milli\meter}. Figure S5C shows the normalized stress relaxation curve. We find the two-term Prony series with $g_1 = 0.07, \tau_1 = \SI{0.08}{\second}, g_2 = 0.05$ and $\tau_2 = \SI{2.05}{\second}$ fits the stress relaxation data well. The total stress relaxation is small ($g_1 + g_2 \approx \SI{10}{\percent}$), indicating a weak viscosity of the hydrogel sample, which only introduces a \textasciitilde \SI{5}{\percent} variation in shear wave speed over the frequency range from \textasciitilde \SI{0.5}{\hertz} ($\tau_2^{-1}$) to \textasciitilde \SI{12.5}{\hertz} ($\tau_1^{-1}$).

While the stress relaxation characterizes the viscoelasticity in the low frequency regime (below \textasciitilde \SI{12.5}{\hertz}), we further measured the surface wave phase velocity up to \SI{800}{\hertz} using our ultrasound elastography system. In this measurement, we relied on a mechanical shaker (SA-JZ002, Shiao, Jiangsu, China) to apply a surface pressure locally to generate harmonic surface waves. The surface waves were acquired by the ultrasound transducer. We then computed the wavelengths of the surface waves to get the phase velocity. As shown in Fig.~S5D, interestingly, we do not observe an increase in the speed, but instead a slight decrease. We attribute this decrease to the slight stiffness gradient (softer at shallower locations) of the hydrogel sample introduced by the fabrication process \citep{ref52}. Despite the slight material heterogeneity, the dispersion relation suggests a weak dependence of the surface wave speed on the frequency, indicating a weak viscosity of the hydrogel in the frequency range of \SIrange{100}{800}{\hertz}.

\subsection*{Supplementary Note 5: Acoustoelastic model for skeletal muscle and the effect of viscoelasticity}
\justify

\textbf{Linear elastic parameters}

To characterize the anisotropy of the skeletal muscle, we measured the shear wave group velocities along different directions. Our main assumption is that the skeletal muscle can be modeled as an incompressible transversely isotropic material due to a preferred direction of the muscle fibers. Such a material has three independent elastic parameters, say $\mu_T$, the transverse shear modulus, $\mu_L$, the longitudinal shear modulus, and $E_L$, the longitudinal Young modulus. We measured the horizontal shear wave speeds $v_x$ in the undeformed material at three different orientations of the fibres with respect to the $x$ axis ($0^\circ, 35^\circ, 90^\circ$, see Figs.~S6A-C) to get $v_{x}^{0^\circ}, v_{x}^{35^\circ}$, and $v_{x}^{90^\circ}$. Then the three elastic parameters can be calculated by the formulas \citep{ref50}
\begin{equation} \label{eq:S15}
\mu_T = \rho(v_{x}^{90^\circ})^2, \quad \mu_L = \rho(v_{x}^{0^\circ})^2, \quad E_L = \frac{4(\rho(v_{x}^{35^\circ})^2 - \mu_L)}{\sin^2(2 \times 35^\circ)} + (4\mu_L - \mu_T).
\end{equation}
Figure S6 shows the statistical results for the shear wave speeds, which clearly points to the mechanical anisotropy of the muscle. From the wave speeds we get $\mu_T \approx \SI{10.7}{\kilo Pa}, \mu_L \approx \SI{22.4}{\kilo Pa}$, and $E_L \approx \SI{40.1}{\kilo Pa}$.

\justify
\textbf{Acoustoelastic model for skeletal muscle}

To model the acoustoelasticity of the skeletal muscle, we take the phenomenological model proposed by Murphy \citep{ref53},
\begin{equation} \label{eq:S16}
W = \frac{\mu_T}{2c_2}[e^{c_2(I_1-3)} - 1] + \frac{E_L + \mu_T - 4\mu_L}{2c_4}[e^{c_4 (\sqrt{I_4}-1)} - 1] + \frac{\mu_T - \mu_L}{2}(2I_4 - I_5 - 1),
\end{equation}
where $c_2$ and $c_4$ are non-dimensional strain-hardening parameters, and the strain invariants are defined as
\begin{equation} \label{eq:S17}
I_1 = \text{tr }\mathbf{C}, \quad I_2 = \tfrac{1}{2}[I_1^2 - \text{tr}(\mathbf{C}^2)], \quad I_4 = \mathbf{M} \cdot (\mathbf{CM}), \quad I_5 = \mathbf{M} \cdot (\mathbf{C}^2\mathbf{M}).
\end{equation}
This model reduces to the neo-Hookean model,
\begin{equation} \label{eq:S18}
W = \mu(I_1 - 3),
\end{equation}
when we take $\mu_T = \mu_L = \frac{1}{3}E_L = \mu$ and $c_2 = 0$.

Inserting \eqref{eq:S16} into \eqref{eq:S2} we obtain the expressions for $\alpha, \beta$, and $\gamma$, which determine the shear wave speed according to Eq.~(1) in the main text. When $\mathbf{M} = (1, 0, 0)$, we find
\begin{align} \label{eq:S19}
\rho v^2 = &\mu_T\lambda^2 \sin^2\theta e^{c_2(I_1-3)} + \lambda^{-1} \cos^2\theta \left[ \mu_T e^{c_2(I_1-3)} + (\mu_T - \mu_L) (2 - 3\lambda^{-1}) \right] \nonumber \\
&+ \lambda^{-1} \cos^2\theta \left[ \frac{(E_L + \mu_T - 4\mu_L)e^{c_4(\lambda-1)}}{2} (1 - \lambda^{-1}) \right],
\end{align}
where $I_1 = \lambda^2 +2\lambda^{-1}$, and $\lambda$ is the stretch ratio along the direction of tension, obtained by solving
\begin{equation} \label{eq:S20}
\sigma_1 = \lambda \left[ \mu_T\lambda e^{c_2(I_1-3)} + \frac{(E_L + \mu_T - 4\mu_L) (\lambda - 1)e^{c_4(\lambda-1)}}{2} + 2(\mu_T - \mu_L) (\lambda - \lambda^3) \right] - \frac{\mu_T}{\lambda} e^{c_2(I_1-3)},
\end{equation}
given the principal stress $\sigma_1$. Figure S1 shows the typical dependence of the wave speed on direction when the material is subject to a uni-axial tension. Inserting $\mu_T = \SI{10.7}{\kilo Pa}, \mu_L = \SI{22.4}{\kilo Pa}$, and $E_L = \SI{40.1}{\kilo Pa}$ into Eq.~\eqref{eq:S19}, and then using this equation to fit $v_x (\theta = 0)$ and $v_z (\theta = \pi/2)$ shown in Fig.~4d, we get $c_2 \approx 3.5$ and $c_4 \approx 8$. The fitting curves are shown in Fig.~4d of the main text.

\justify

\textbf{Viscoelasticity of the skeletal muscle and its effect on shear wave propagation}

The dispersion relation of the Rayleigh surface wave in the muscle sample was measured using the same setup as described in Note 4. Figure S6E shows the surface wave speeds measured along the muscle fiber. We fit the dispersion relation with a one-term Prony series, to get $g_1 \approx 0.79$ and $\tau_1 \approx \SI{0.49}{\milli\second}$.

To evaluate the effect of the viscoelasticity on the acoustoelastic imaging, we use the acousto-visco-elastic model recently proposed by Berjamin and de Pascalis \citep{ref45}. For simplicity, we consider the quasi-linear viscoelasticity (QLV) theory with the neo-Hookean model (Eq.~\eqref{eq:S18}) and a one-term Prony series. According to \citep{ref45}, the shear wave speed $v_x$ is a function of the frequency $f$,
\begin{equation} \label{eq:S21}
v_x = \sqrt{\frac{2(1 + D^2)}{1 + \sqrt{1 + D^2}}} \sqrt{\frac{|\text{Re} \mu_x|}{\rho}}
\end{equation}
where
\begin{gather}
D = D_0 \frac{2\Omega\Omega_0}{\Omega^2 + \Omega_0^2}, \quad \Omega = 2\pi f \tau_1, \label{eq:S22a} \\
D_0 = \frac{g_1}{2\Omega_0} \frac{\bar{\mu}^v_x}{\bar{\mu}^v_x + (1 - g_1) [\bar{T}^e_d]_{11}}, \quad \Omega_0^2 = (1 - g_1) \frac{\bar{\mu}^v_x + [\bar{T}^e_d]_{11}}{\bar{\mu}^v_x + (1 - g_1) [\bar{T}^e_d]_{11}}, \label{eq:S22b}
\end{gather}
and
\begin{equation} \label{eq:S23}
\mu_x = (1 - g_1) [\bar{T}^e_d]_{11} + \left(1 - \frac{g_1}{1 + i\omega\tau_1}\right)\bar{\mu}^v_x.
\end{equation}
In \eqref{eq:S22b} and \eqref{eq:S23}, $i = \sqrt{-1}$,
\begin{equation} \label{eq:S24}
[\bar{T}^e_d]_{11} = \mu(\lambda^2 - I_1/3), \quad \bar{\mu}^v_x = \mu I_1/3.
\end{equation}
and $\lambda$ is the stretch ratio, which can be determined from the stress $\sigma_1$ by solving the cubic
\begin{equation} \label{eq:S25}
\lambda^3 - \frac{\sigma_1}{\mu_\infty}\lambda - 1 = 0,
\end{equation}
where $\mu_\infty = \mu(\infty) = (1 - g_1)\mu_0$ is the long-term shear modulus and $I_1 = \lambda^2 + 2\lambda^{-1}$. To get $v_z$, we follow the same procedure, replacing $\lambda$ with $\lambda^{-1/2}$ in \eqref{eq:S24}.

In Fig.~S7, we plot the dispersion relations of $v_x$ and $v_z$ with $\mu_\infty = \SI{8.4}{\kilo Pa}$, $g_1 = 0.79$ and $\tau_1 = \SI{0.49}{\milli\second}$. Then we use Eq.~(2) in the main text to derive the stress $\sigma_1$. As shown in Fig.~S7B, the stress is underestimated when the viscoelasticity comes into play.

\begin{figure}[p]
    \centering
    \includegraphics[width=0.8\linewidth]{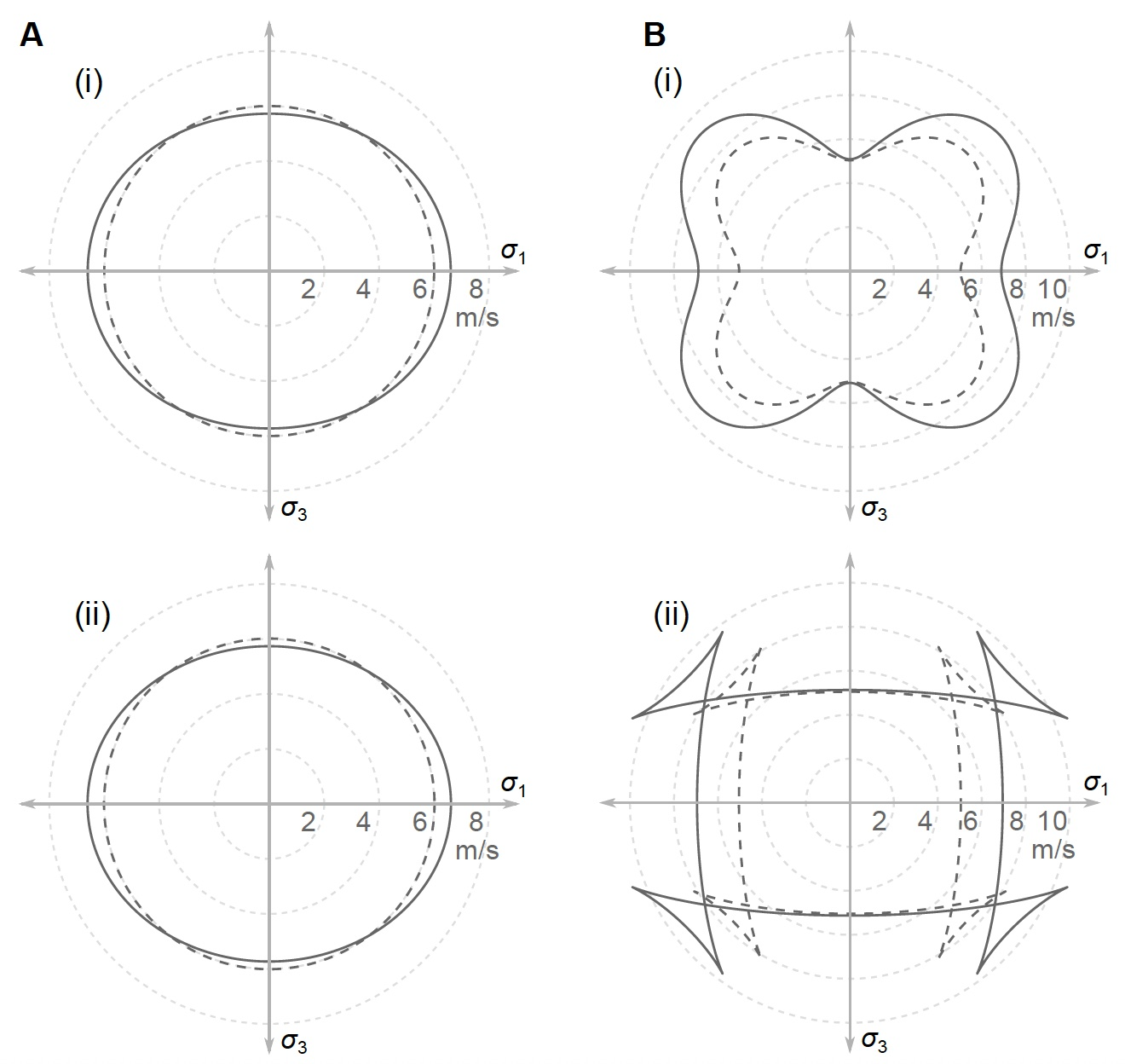}
    \caption{Effect of the uniaxial stress on the shear wave speeds. (A) neo-Hookean material with shear modulus $\mu = \SI{36}{\kilo Pa}$, subject to uniaxial stress $\sigma_1 = 0.3\mu$. (B) Transversely isotropic material with material parameters $\mu_T = \SI{9}{\kilo Pa}$, $\mu_L = \SI{25}{\kilo Pa}$, $E_L = \SI{216}{\kilo Pa}$, $c_1 = 1$, $c_2 = 10$, and $\sigma_1/E_L = 0.1$. The fiber direction is aligned with $x_1$. (i) and (ii) depict phase and group speeds, respectively. Solid lines: prestressed. Dashed lines: stress-free.}
    \label{fig:S1}
\end{figure}

\begin{figure}[p]
    \centering
    \includegraphics[width=\linewidth]{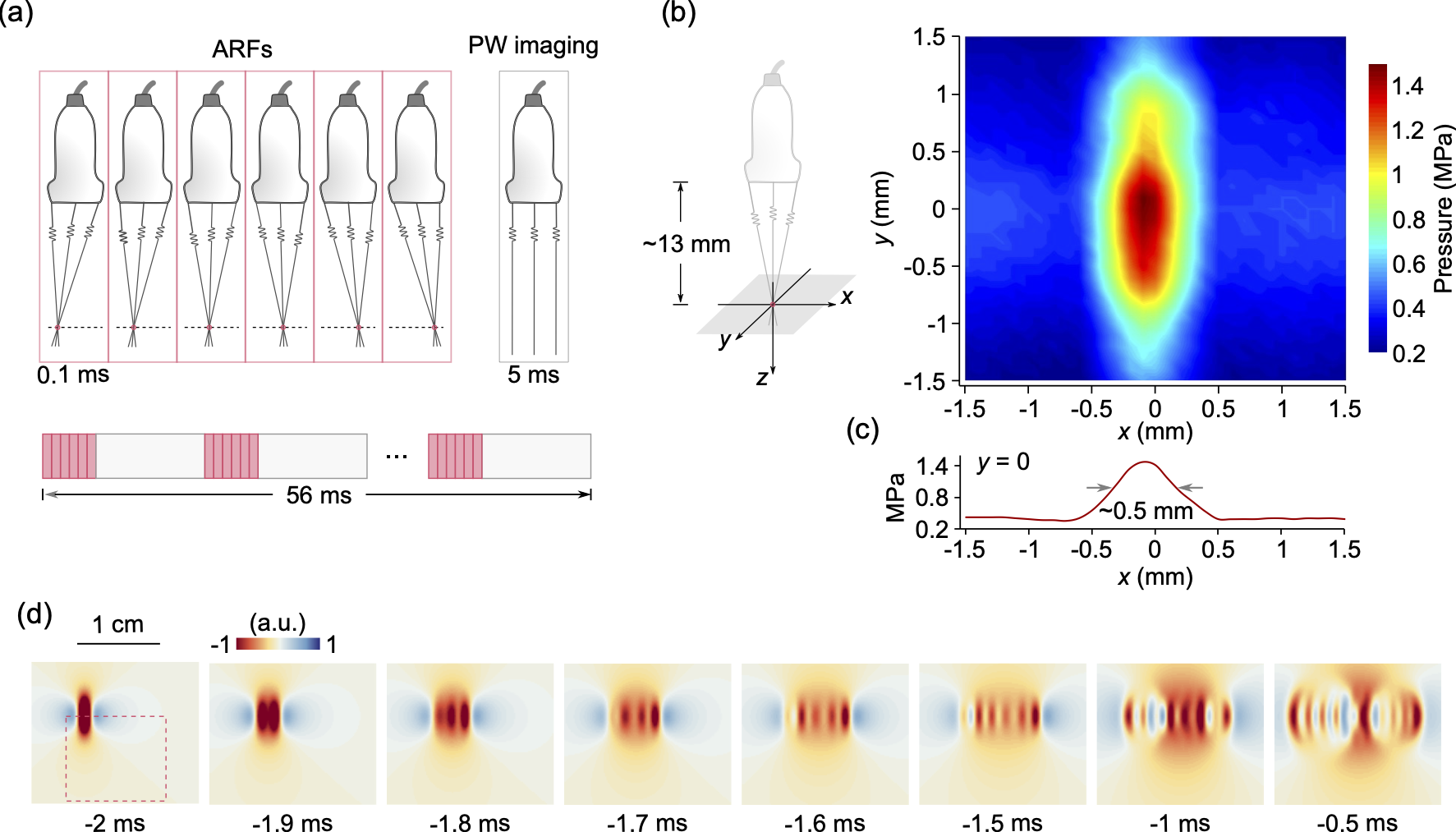}
    \caption{Imaging protocol and finite element simulation of shear wave excitation. (a) Imaging protocol. Six ARFs are applied by successively focusing the ultrasound beam along horizontal direction. The duration of each ARF is \textasciitilde \SI{0.1}{\milli\second}. After the excitation (\textasciitilde \SI{0.6}{\milli\second}), the transducer is switched to perform plane wave (PW) imaging (unfocused beam, duration \SI{5}{\milli\second}) at a frame rate of \SI{10}{\kilo\hertz}. Ten successive measurements (\textasciitilde \SI{56}{\milli\second}) are performed and then the average of the measurements is taken to improve the signal-to-noise ratio. (b) Acoustic pressure of the focused ultrasound beam measured within the focal plane (\textasciitilde \SI{13}{\milli\meter} away from the transducer). (c) Distribution of the pressure along $x$ axis. Half width at half maximum (HWHM) is approximately \textasciitilde \SI{0.25}{\milli\meter}, in agreement with the ultrasound wavelength \textasciitilde \SI{0.23}{\milli\meter}. (d) Finite element simulations showing the six ARFs successively applied to excite the shear waves. The time when the PW imaging starts is set to be 0. The dashed square shows the region of interest where the wave propagation is measured by the PW imaging.}
    \label{fig:S2}
\end{figure}

\begin{figure}[p]
    \centering
    \includegraphics[width=\linewidth]{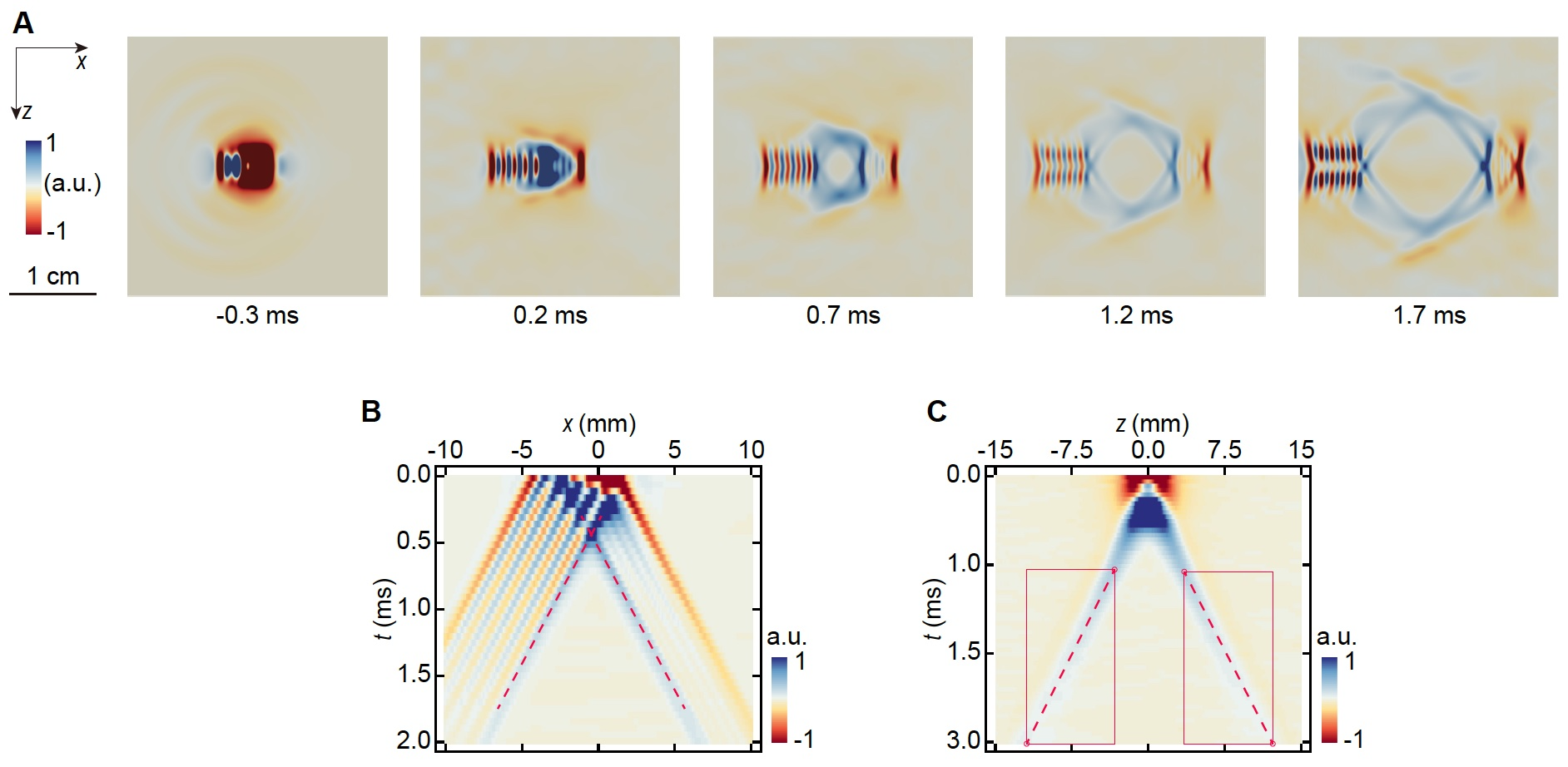}
    \caption{Finite element simulation of the shear wave excitation by programmed acoustic radiation forces in anisotropic materials. (a) The snapshots of the shear wave propagation, which suggest the SV shear waves are primarily excited. The maps depict the vertical particle velocity fields. (b) and (c) Spatiotemporal data for the horizontal and vertical waves, respectively. The speeds measured along the two directions are identical, \textasciitilde \SI{4.7}{\meter\per\second} ($\sqrt{\mu_L/\rho}$), indicating the SV shear waves are measured in both directions. The material is incompressible transversely isotropic. The fiber direction is aligned with $x$. The material parameters used in the simulation are $\mu_T = \SI{10.7}{\kilo Pa}$, $\mu_L = \SI{22.4}{\kilo Pa}$, $E_L = \SI{40.1}{\kilo Pa}$, and $\rho = \SI{1000}{\kilo\gram\per\meter\cubed}$ (see Note 5 for definitions of the material parameters).}
    \label{fig:S3}
\end{figure}

\begin{figure}[p]
    \centering
    \includegraphics[width=\linewidth]{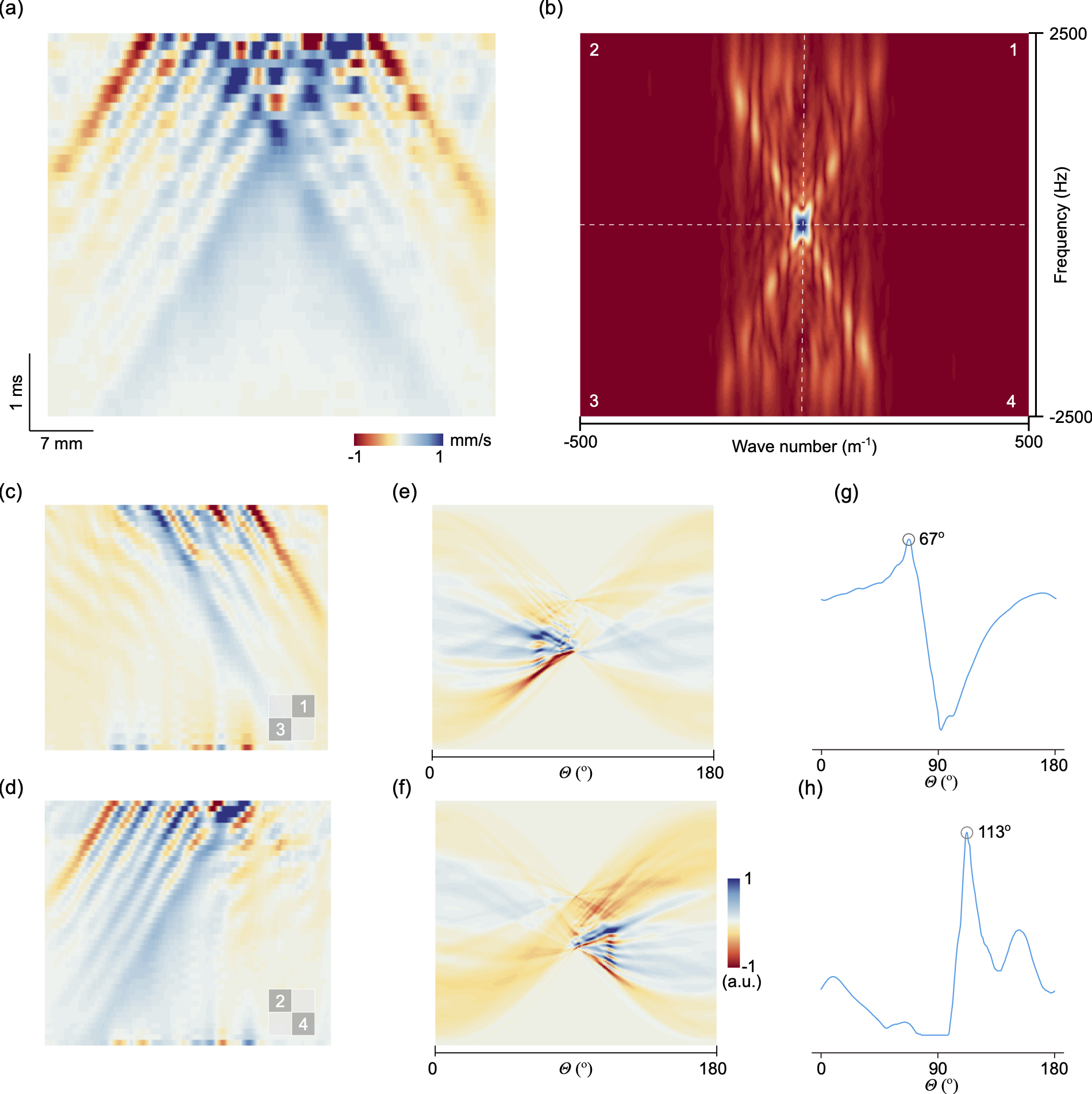}
    \caption{Measurement of the lateral shear wave speed $v_x$. (a) Spatiotemporal map of the shear waves propagating along the horizontal direction ($x$ axis). (b) Fourier transformation of the spatiotemporal data. (c) Inverse Fourier transformation of the data in the first and third quadrants. The right-to-left (RL) waves have been filtered out in this map. (d) Inverse Fourier transformation of the data in the second and fourth quadrants. The left-to-right (LR) waves have been filtered out in this map. (e) and (f) The Radon transformations of (c) and (d). Then We sum the absolute values of the data points in (e) and (f) along each column to get the solid lines in (g) and (h), respectively. The peaks identified on the lines give the optimal group velocities of the LR and RL waves.}
    \label{fig:S4}
\end{figure}

\begin{figure}[p]
    \centering
    \includegraphics[width=\linewidth]{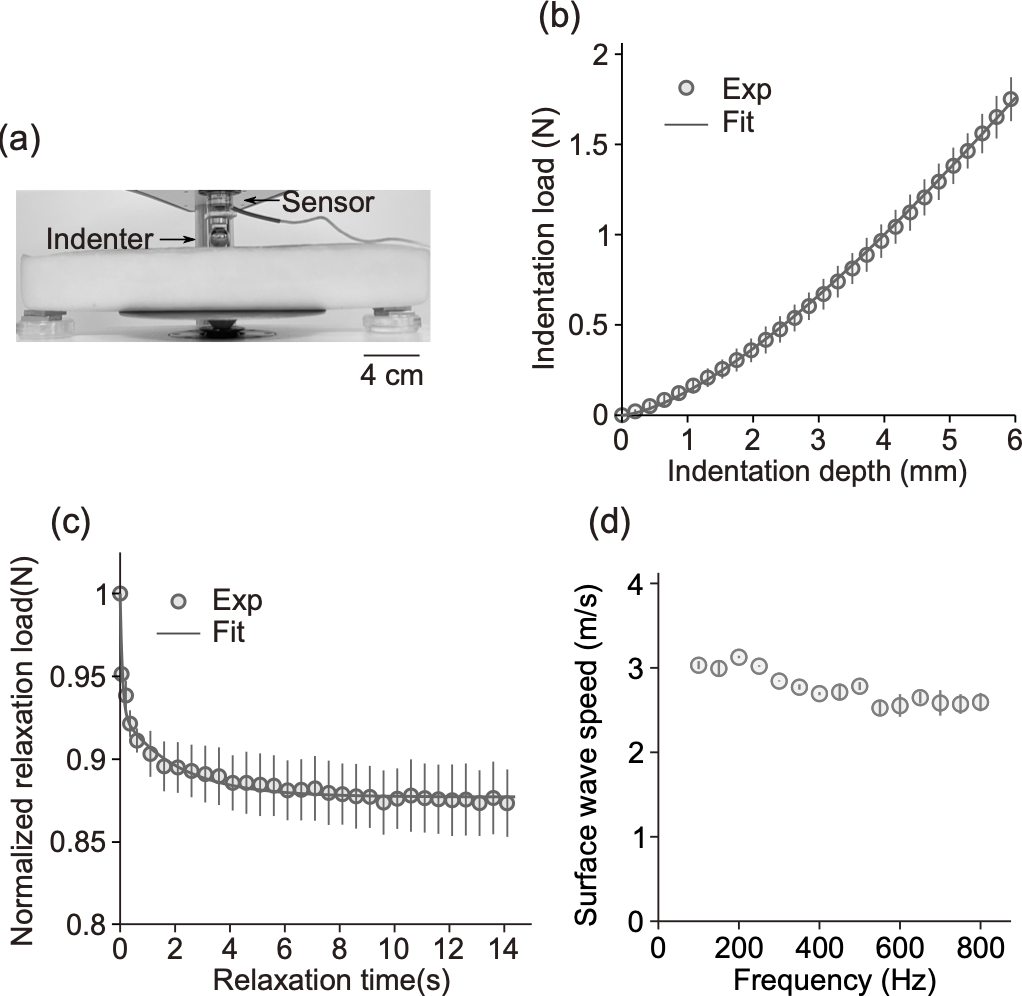}
    \caption{Mechanical characterization of the hydrogel phantom at rest. (a) Photography showing the indentation tests on the hydrogel phantom. (b) Load-displacement curve of the indentation experiments obtained from the loading process with a low loading rate (\textasciitilde \SI{0.1}{\milli\meter\per\second}). Error bar, standard deviations over five measurements. (c) Normalized stress relaxation curve. Error bar: standard deviations over ten measurements. (d) Phase velocity of the surface waves.}
    \label{fig:S5}
\end{figure}

\begin{figure}[p]
    \centering
    \includegraphics[width=\linewidth]{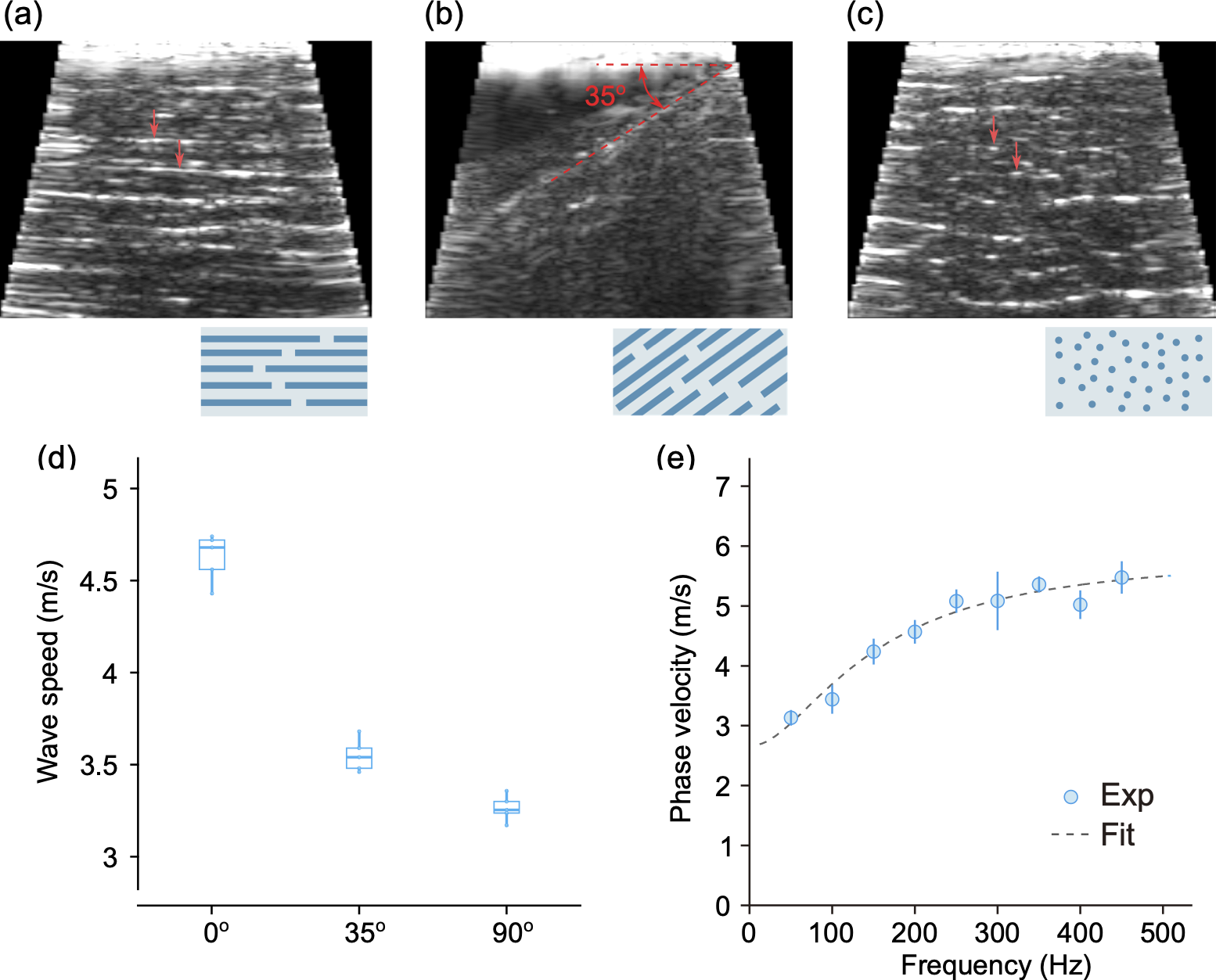}
    \caption{Mechanical characterization of the skeletal muscle at rest. (a)-(c) Grayscale ultrasound images of the skeletal muscle. Red arrows in (a) and (c) indicate some of the parallel muscle fibers. For (b) the sample is tilted at \textasciitilde $35^\circ$. The schematics underneath each image show the orientations of the muscle fibers. For all three cases, the horizontal shear wave group velocities $v_x$ are measured. Therefore, the angles between the shear wave propagation direction and muscle fibers are (a) $0^\circ$, (b) $35^\circ$, and (c) $90^\circ$. (d) Statistical results (five independent measurements) for the horizontal shear wave group velocities. (e) Dispersion relation of the surface waves ($0^\circ$). Markers, experiments. Dashed line, fitting curve obtained using one-term Prony series with $g_1 = 0.79$ and $\tau_1 = \SI{0.49}{\milli\second}$.}
    \label{fig:S6}
\end{figure}

\begin{figure}[p]
    \centering
    \includegraphics[width=0.8\linewidth]{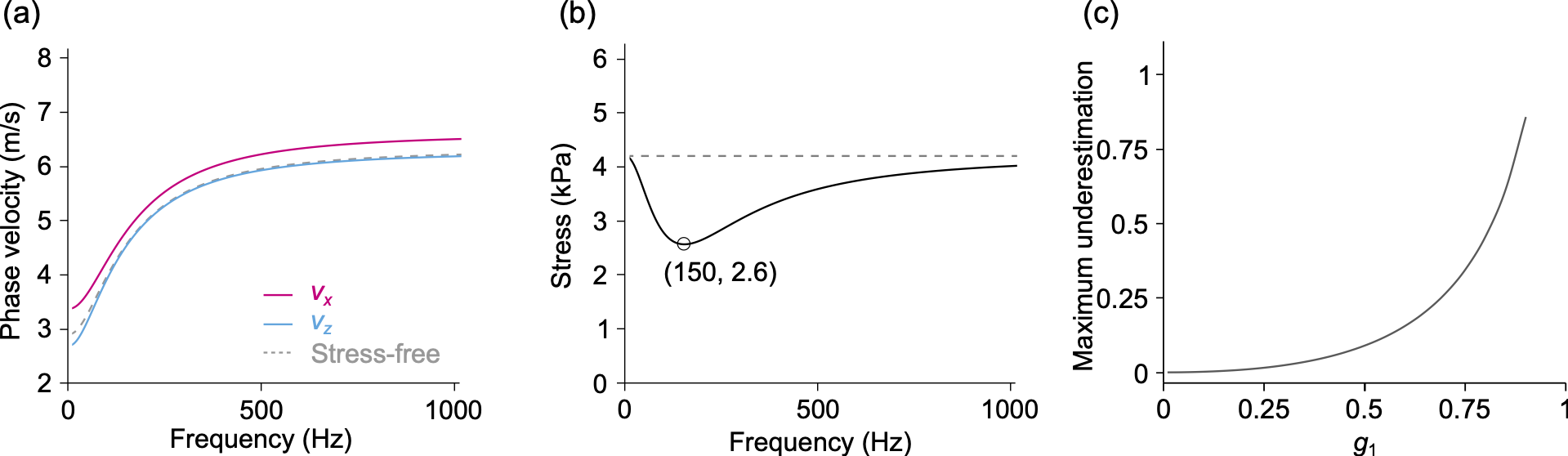}
    \caption{Effect of viscoelasticity on the acoustoelastic imaging. (a) Dispersion relations of $v_x$ and $v_z$ when a tensile stress $\sigma_1 = \SI{4.2}{\kilo Pa}$ is applied. The dashed curve is the dispersion relation in the stress free state. The Quasi-Linear Viscoelastic material model used to produce this figure relies on the neo-Hookean model with $\mu_0 = \SI{40}{\kilo Pa}$ and the one-term Prony series with $g_1 = 0.79$ and $\tau = \SI{0.49}{\milli\second}$. (b) The stresses derived from $v_x$ and $v_z$ at different frequencies. The minimum stress is \SI{2.6}{\kilo Pa}, indicating an underestimation of \textasciitilde \SI{38}{\percent}. (c) The underestimation of the stress as a function of $g_1$.}
    \label{fig:S7}
\end{figure}

\begin{figure}[p]
    \centering
    \includegraphics[width=0.7\linewidth]{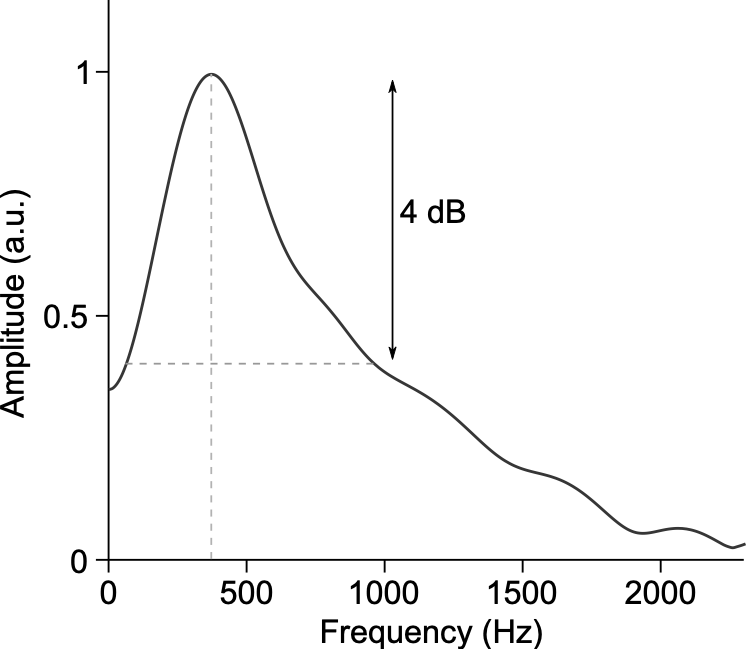}
    \caption{A representative spectrum of the shear waves in the muscle sample. The central frequency is about \SI{380}{\hertz} and the \SI{4}{dB} bandwidth is about from \SIrange{100}{1000}{\hertz}.}
    \label{fig:S8}
\end{figure}

\clearpage
\section*{Supplementary Movie Legends}
\begin{itemize}
    \item \textbf{Movie S1} (separate file). Finite element simulations of the shear waves generated by (A) the programmed acoustic radiation force (ARF) and (B) a single ARF.
    \item \textbf{Movie S2} (separate file). Experimental measurements of the shear waves generated by (A) the programmed acoustic radiation force (ARF) and (B) a single ARF.
\end{itemize}


\clearpage
\begin{thebibliography}{53}
\bibitem{ref1} N. L. Nerurkar, C. Lee, L. Mahadevan, C. J. Tabin, Nature \textbf{565}, 480 (2019).
\bibitem{ref2} M. Gómez-González, E. Latorre, M. Arroyo, X. Trepat, Nat. Rev. Phys. \textbf{2}, 300 (2020).
\bibitem{ref3} K. H. Vining, D. J. Mooney, Nat. Rev. Mol. Cell Biol. \textbf{18}, 728 (2017).
\bibitem{ref4} H. Ucar, et al., Nature \textbf{600}, 686–689 (2021).
\bibitem{ref5} T. Tallinen, et al., Nat. Phys. \textbf{12}, 588 (2016).
\bibitem{ref6} J.-H. Lee, H. S. Park, D. P. Holmes, Phys. Rev. Lett. \textbf{127}, 138102 (2021).
\bibitem{ref7} B. Li, F. Jia, Y.-P. Cao, X.-Q. Feng, H. Gao, Phys. Rev. Lett. \textbf{106}, 234301 (2011).
\bibitem{ref8} S. K. Powers, M. J. Jackson, Physiol. Rev. \textbf{88}, 1243 (2008).
\bibitem{ref9} J. A. Martin, et al., Nat. Commun. \textbf{9}, 1 (2018).
\bibitem{ref10} J. Lee, et al., Nat. Electron. \textbf{4}, 291 (2021).
\bibitem{ref11} J. Y. Sun, et al., Nature \textbf{489}, 133 (2012).
\bibitem{ref12} J. Kim, G. Zhang, M. Shi, Z. Suo, Science \textbf{374}, 212 (2021).
\bibitem{ref13} N. Matsuhisa, et al., Nature \textbf{600}, 246 (2021).
\bibitem{ref14} G. H. Lee, et al., Nat. Rev. Mater. \textbf{5}, 149 (2020).
\bibitem{ref15} R. Bai, J. Yang, Z. Suo, Eur. J. Mech. A/Solids \textbf{74}, 337 (2019).
\bibitem{ref16} J. A. Rogers, T. Someya, Y. Huang, Science \textbf{327}, 1603 (2010).
\bibitem{ref17} Y. Wang, et al., Sci. Adv. \textbf{6}, 7043 (2020).
\bibitem{ref18} G. S. Schajer, Exp. Mech. \textbf{50}, 245 (2009).
\bibitem{ref19} N. S. Rossini, M. Dassisti, K. Y. Benyounis, A. G. Olabi, Mater. Des. \textbf{35}, 572 (2012).
\bibitem{ref20} C. O. Ruud, NDT E Int. \textbf{15}, 15 (1982).
\bibitem{ref21} D. S. Hughes, J. L. Kelly, Phys. Rev. \textbf{92}, 1145 (1953).
\bibitem{ref22} A. N. Guz, F. G. Makhort, Int. Appl. Mech. \textbf{36}, 1119 (2000).
\bibitem{ref23} F. Shi, J. E. Michaels, S. J. Lee, J. Acoust. Soc. Am. \textbf{133}, 677 (2013).
\bibitem{ref24} J.-L. Gennisson, et al., J. Acoust. Soc. Am. \textbf{122}, 3211 (2008).
\bibitem{ref25} G.-Y. Li, A. Gower, M. Destrade, J. Acoust. Soc. Am. \textbf{148}, 3963 (2020).
\bibitem{ref26} Y. Jiang, et al., Med. Image Anal. \textbf{20}, 97 (2015).
\bibitem{ref27} C. Creton, C. Matteo, Rep. Prog. Phys. \textbf{79}, 046601 (2016).
\bibitem{ref28} M. Shams, M. Destrade, R. W. Ogden, Wave Motion \textbf{48}, 552 (2011).
\bibitem{ref29} A. L. Gower, T. Shearer, P. Ciarletta, The Quarterly Journal of Mechanics and Applied Mathematics \textbf{70}, 455 (2017).
\bibitem{ref30} M. Destrade, M. D. Gilchrist, R. W. Ogden, The Journal of the Acoustical Society of America \textbf{127}, 2103 (2010).
\bibitem{ref31} C. Deroy, M. Destrade, A. Mc Alinden, A. Ní Annaidh, Skin Research and Technology \textbf{23}, 326 (2017).
\bibitem{ref32} I. Hariton, G. Debotton, T. C. Gasser, G. A. Holzapfel, Biomechanics and modeling in mechanobiology \textbf{6}, 163 (2007).
\bibitem{ref33} G. W. Jones, S. J. Chapman, Siam review \textbf{54}, 52 (2012).
\bibitem{ref34} N. C. Rouze, A. Caenen, K. R. Nightingale, Physics in Medicine \& Biology \textbf{67}, 095015 (2022).
\bibitem{ref35} J. Bercoff, M. Tanter, M. Fink, IEEE Trans. Ultrason. Ferroelectr. Freq. Control \textbf{51}, 396 (2004).
\bibitem{ref36} A. P. Sarvazyan, O. V. Rudenko, S. D. Swanson, J. B. Fowlkes, S. Y. Emelianov, Ultrasound Med. Biol. \textbf{24}, 1419 (1998).
\bibitem{ref37} S. Catheline, N. Benech, J. Acoust. Soc. Am. \textbf{137}, EL200 (2015).
\bibitem{ref38} L. Sandrin, M. Tanter, J.-L. Gennisson, S. Catheline, M. Fink, IEEE transactions on ultrasonics, ferroelectrics, and frequency control \textbf{49}, 436 (2002).
\bibitem{ref39} L. Sandrin, D. Cassereau, M. Fink, The Journal of the Acoustical Society of America \textbf{115}, 73 (2004).
\bibitem{ref40} M. Tanter, M. Fink, IEEE Trans. Ultrason. Ferroelectr. Freq. Control \textbf{61}, 102 (2014).
\bibitem{ref41} N. Rouze, M. Wang, M. Palmeri, K. Nightingale, IEEE Trans. Ultrason. Ferroelectr. Freq. Control \textbf{57}, 2662 (2010).
\bibitem{ref42} J. P. Kerris, A. C. Betik, J. Li, G. K. McConell, J. Appl. Physiol. \textbf{126}, 239 (2019).
\bibitem{ref43} B. Calvo, et al., J. Biomech. \textbf{43}, 318 (2010).
\bibitem{ref44} J. P. Remeniéras, et al., Phys. Med. Biol. \textbf{66}, 145009 (2021).
\bibitem{ref45} H. Berjamin, R. De Pascalis, International Journal of Solids and Structures \textbf{241}, 111529 (2022).
\bibitem{ref46} G. Montaldo, M. Tanter, J. Bercoff, N. Benech, M. Fink, IEEE transactions on ultrasonics, ferroelectrics, and frequency control \textbf{56}, 489 (2009).
\bibitem{ref47} T. Loupas, J. Powers, R. W. Gill, IEEE transactions on ultrasonics, ferroelectrics, and frequency control \textbf{42}, 672 (1995).
\bibitem{ref48} T. J. Hall, M. Bilgen, M. F. Insana, T. A. Krouskop, IEEE Trans. Ultrason. Ferroelectr. Freq. Control \textbf{44}, 1355 (1997).
\bibitem{ref49} R. W. Ogden, Waves in Nonlinear Pre-Stressed Materials, M. Destrade, G. Saccomandi, eds. (Springer, Vienna, 2007), vol. 495, pp. 1–26.
\bibitem{ref50} G.-Y. Li, Y. Zheng, Y. Liu, M. Destrade, Y. Cao, J. Mech. Phys. Solids \textbf{96}, 388 (2016).
\bibitem{ref51} M. Destrade, M. D. Gilchrist, G. Saccomandi, J. Acoust. Soc. Am. \textbf{127}, 2759 (2010).
\bibitem{ref52} G. Y. Li, et al., Philos. Trans. Royal Soc. A \textbf{377}, 20180075 (2019).
\bibitem{ref53} J. G. Murphy, Eur. J. Mech. A/Solids \textbf{42}, 90 (2013).

\end{thebibliography}
\end{document}